\documentstyle[sao2,fleqn,psfig]{article}
\newcommand{\ch}{CH\,Cyg}
\newcommand{\mwc}{MWC\,560}
\newcommand{\kms}{km\,s$^{-1}$}
\newcommand{\si}{$\sim$}
\newcommand{\msy}{$M_\odot\,{\rm y}^{-1}$}

\begin{document}
\thispagestyle{empty}

\title{Magnetic propeller in symbiotic stars}
\author{Alexander Panferov \inst{a} \and
Maciej Miko{\l}ajewski \inst{b}}
\institute{
Special Astrophysical Observatory,
Nizhnij Arkhyz,
369167, Russia
e-mail: panf@sao.ru \and
Centre for Astronomy,
Nicolaus Copernicus University,
Gagarina 11,
PL-87100 Toru\'n, Poland
e-mail: Maciej.Mikolajewski@astri.uni.torun.pl}

\maketitle

   \begin{abstract} Rapidly spinning magnetic white dwarfs in
symbiotic stars may pass through the propeller stage. It is believed
that a magnetic propeller acts in two such stars \ch\ and
   \mwc. We review a diversity of manifestations of the propeller
there. In these systems in a quiescent state the accretion onto a
white dwarf from the strong enough wind of a companion star is
suppressed by the magnetic field, and the hot component luminosity is
low. Since the gas stored in the envelope eventually settles to the
corotation radius the star-propeller reveals itself by long-time
outbursts, short-time brightness dips, strong amplitude flickering,
ejection of a powerful wind and jets. The envelope appears to be 
stratified and inhomogeneous because of the
shock pumping of the propeller. At
this stage, at which catastrophic transitions to the accretion stage
and back are possible, the release of energy and angular momentum of
the white dwarf is maximum. Using this property, we estimate the
magnetic field strength of the white dwarfs in \ch\ and \mwc\ to be
of order $10^8$ G.

   The most spectacular property of these stars --- jets --- is
inherently connected with the propeller. This suggests hydromagnetic
models of jet acceleration.
       
\end{abstract}

\section{Introduction}

   In the course of spin deceleration a neutron star having a strong
magnetic field may pass through an evolution stage of the propeller
before accreting gas from the surroundings and appearing as 
a bright X-ray source (Schwartzman 1971, Illarionov \& Sunyaev 1975).
However, from observations propellers are not known reliably.
The transfer from the
propeller stage to accretion one is abrupt. Such transfers are
suggested only in the low mass X-ray binaries Aql\,X1 and
SAX\,J1808.4-3658 (Zhang, Yu \& Zhang 1998, Stella et al. 2000). There
are difficulties in recognition of stars at the propeller stage because
of their faintness.

   Stars-propellers escape  apparent identification and may appear
unusual. So, the whole class of cataclysmic variables (CVs) of
nova-like type and particularly that of SW\,Sex type, having some
anomalies, have been suggested by Horne (1999) as stars with disk-anchored
propellers. But the only reliable example of the propeller is fascinating
AE\,Aqr (Wynn, King \& Horne 1997), where the rapidly spinning magnetic
white dwarf (WD) operates as a propeller. Possibly, propellers work in
the striking symbiotic stars (SSs) \ch\ and \mwc\ too.
The propeller in \ch\ was suggested
long ago (Miko{\l}ajewski \& Miko{\l}ajewska 1988), but this
is out of view of
theorists until, in spite of scarcity of data on propellers.
There are other SSs similar in some respects to \ch\ and \mwc.
Therefore a magnetic propeller may be an widespread phenomenon amongst
SSs. Here we study the observation
appearances of a propeller by the examples of \ch\ and \mwc.

   In Sect.~3 we discuss the observational properties of \ch\ and \mwc\
in a framework of the magnetic propeller model, outlined in Sect.~2. The
knowledge of observational signs of a propeller would allow it to be
discriminated one amongst another object and we summarize them in
Sect.~4.

\section{A propeller model}
A magnetic propeller model has been discussed in details everywhere 
(Lipunov 1992 and references therein). We use one of the variants of the
model.

Let a WD of radius $R_{\rm x}$ has a surface dipolar magnetic field
$B_{\rm s}$ and a rotation period $P_{\rm x}$. The gas captured
gravitationally
by the WD settles on the star magnetosphere. Below the magnetosphere
boundary the infalling gas is controlled by the magnetic
field and corotates with the star. The gas accretes onto the
surface of the star when the magnetosphere radius $R_{\rm m}$ is
smaller than the corotation radius, at which the keplerian velocity
equals the velocity of corotation with the star,
$R_{\rm cor}=\left(\frac{G\,M_{\rm x}}{\omega_{\rm x}^2}\right)^{1/3}$,
where $G$ is the gravitation constant, $M_{\rm x}$ mass of the WD
and $\omega_{\rm x}=2\,\pi/P_{\rm x}$ its angular velocity. The
accretion occurs because
the gravitational acceleration dominates over the centrifugal
acceleration. The reverse takes place and the infalling gas is expelled
from the magnetosphere when $R_{\rm m} > R_{\rm cor}$ --- the rotating
magnetic field acts as a propeller.
Due to coexistence of the inflow and the
outflow and hence their counteraction each other
around the star-propeller, the accretion
structure may be complex there. This is not well understood yet. It
is expected that the expelled mass does not leave the star and the envelope
builds up around the magnetosphere. In a steady state the
magnetic pressure $p_{\rm m}$ at the upper boundary of the magnetosphere
balances the gas pressure $p_{\rm g}$ just above this
boundary:
\begin{equation}
\label{bal}
p_{\rm m}=\frac{B_{\rm s}^2}{8\,\pi} 
\left(\frac{R_{\rm x}}{R_{\rm m}}\right)^6=p_{\rm g}.
\end{equation}

While the mass accumulates, the envelope contracts the
magnetosphere until the onset of the accretion phase
($R_{\rm m} < R_{\rm cor}$).
The evolution of the envelope is connected with
behavior of the magnetosphere and a simulation of this in
comparison with the observations would give an insight into
how a propeller operates.

The interaction between the envelope and rotating magnetic field
of the star results in a development of a shear layer of width
$\delta$  comparable with the free fall length
(Wang \& Robertson 1985): 
$\delta \sim \left(\frac{\eta v_{\rm esc}}
{\omega_{\rm x} R_{\rm m}}\right)^2 R_{\rm m}$,
where $v_{\rm esc}=\sqrt{2\,M\,G/R_{\rm m}}$ is the escape
velocity at $R_{\rm m}$, $\eta$ {\em in situ} the effectiveness of
damping of the convective motions in the shear layer by turbulent
viscosity and is of order 1. This width is
determined by convection, i.e. by the gravitation effect. In the subsonic
case ($\omega_{\rm x} R_{\rm m} \la c_{\rm s}$), which takes place
when the magnetosphere radius is $R_{\rm m} \la R_{\rm cor}$,
the form $\delta \sim R_{\rm m}$ is more appropriate for
the shear layer width.
The matter of the envelope penetrates into the shear layer via
instabilities. The magnetic field mixes the matter and forces
it to corotate. These processes determine the radius and state of the
envelope.

In the shear layer the star-propeller transfers outside the angular
moment and releases energy of rotation. The equality of the 
Alfvenic and the corotation velocities at the magnetosphere boundary,
$B^2/ 4\,\pi\,\rho = (\omega\, R_{\rm m})^2$, is expected
in a steady state on average (Wang \& Robertson 1985).
Then the energy release of the propeller is:
\begin{equation}
\label{lprop}
L_{\rm p} \sim \frac{B^2}{8\,\pi\, P_{\rm x}} \times 4\,\pi
R_{\rm m}^2 \delta.
\end{equation}

\section{Manifestations of the propeller in \protect\\ \ch\ and \mwc}
\ch\ and \mwc\ are symbiotic long-period binaries undergoing occasional
2 -- 3 mag outbursts, as do classical SSs, but moreover showing
a significant mass loss and jet ejection. Their hot continuum radiation
originates in the envelope around the WD. The envelope is observed
in bright states, therefore it exists at least some yeas.
The conventional models of symbiotic stars
(Miko{\l}ajewska \& Kenyon 1992) fail to account for activity in
these stars and, in particular, for their high-velocity lumpy outflows and
jets. The parameters of these binaries are given in Table~1.

\begin{table}
\caption{The binaries parameters.}
\begin{center}
\begin{tabular}{lrr}\hline
                                         & \ch     & \mwc     \\
\hline
$\gamma$ (\kms)                          &-57.7    & 35       \\
$K_{\rm g}$ (\kms)                       &  4.9    & $<2$     \\
$e$                                      &  0.47   & $\approx 0.3$     \\
$M_{\rm g}$ (M$_{\odot}$)                &3.5, M6 -- 7\,III&M4 -- 5\,III\\
$M_{\rm g}/M_{\rm wd}$                   &  3.5    &          \\
$P_{\rm orb}$ (days)                     & 5700    & 1930     \\
$w_{\rm g}$ ($^\circ$)                   &  142    &          \\
$i_{\rm orb}$ ($^\circ$)                 &   90    & $<10$    \\
$a_{\rm orb}$ (cm)             & $1.4\,10^{14}$    & $\approx 6\, 10^{13}$ \\
$\dot{M}_{\rm w}$ (\msy)            & $\sim 10^{-7}$ & $\sim 10^{-6}$\\
$V_{\rm w}$ (\kms)                       & 30      &          \\
\hline 
$P_{\rm x}$ (sec)                        & 500     & 1320     \\
$R_{\rm co}$ (cm)                        & $9.4\,10^9$ & $<2\,10^{10}$ \\
$V_{\rm eject}$ (\kms)                   & 3000    & 7000     \\
\hline
\end{tabular}
\end{center}
\label{orbpar}
\end{table}

\subsection{\ch}
   The semi-regular variable \ch\ entered in period of
activity in 1963 after almost a century of its observations in 
quiescence (Fig.~1). The outbursts are characterized by appearance
of a hot blue continuum, strong emission lines of HI,
HeI, FeII and [FeII] and short time photometric variability
(flickering), similar to that of CVs. In the outbursts 
coherent oscillations of light with periods of 500 and 2000 -- 3000 sec
were
observed (Miko{\l}ajewski et al. 1990, Hoard 1993).
The duration of
the outbursts is from one to several years. The most spectacular phenomenon
was a sudden drop of brightness
in 1984 with simultaneous ejection of jets with a velocity of about
1000 km/s. Intriguingly, this resembles outbursts in X-ray transients
ejecting jets but on much longer timescales.
In the outbursts the kinematics and the ionization of the gas   evidence
a high inhomogeneity and a shock pumping of radiation of
the envelope around the WD (Faraggiana \& Hack 1971,
Persic, Hack \& Selvelli 1984). The 
simultaneous fall and outflow of gas are observed
in the envelope. The temperature
of the envelope changes in the course of an outburst in a range from
7000 to 15000\,K (Mikolajewska, Selvelli \& Hack 1988).

\begin{figure}
\centerline{\psfig{figure=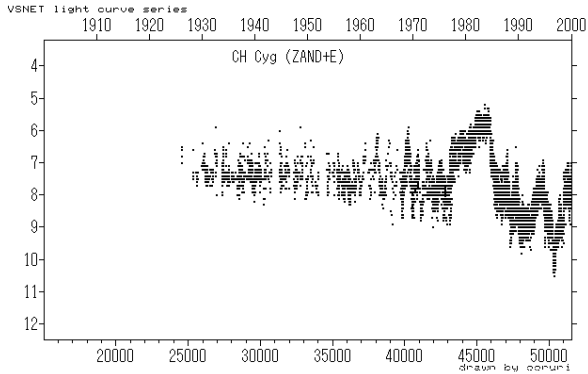,height=5cm}}
\caption{The long-term light curve of \ch\ in V band (from VSNET,
http://www.kusastro.kyoto-u.ac.jp/vsnet).}
\end{figure}

\begin{figure}
\centerline{\psfig{figure=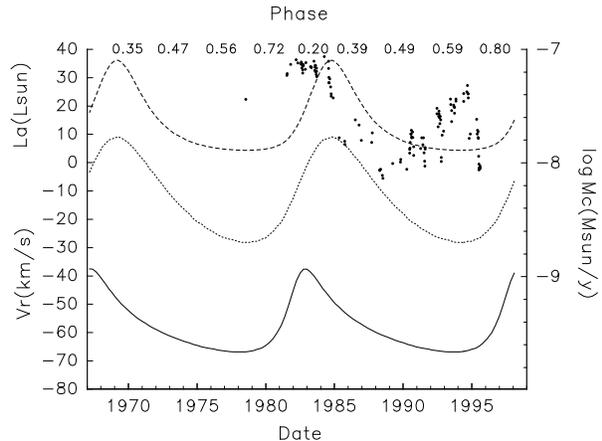,height=6cm}}
\caption{CH\,Cyg model orbital dependences of the radial velocities 
of the WD (solid line), the capture mass rate
(doted line) and luminosity of the accretion onto
the WD surface (dashed line). Compare these with the behaviour of observed
U brightness (dots), which is scaled by expression ${\rm U'=-U/6-6.2}$
to fit right axis.}
\label{orbita}
\end{figure}

The activity of the compact companion in the binary \ch\ is powered by gas
accretion from the giant's wind. With the adopted parameters of
the binary orbit and wind, the accretion mass rate is
$\dot M_{\rm a}= \dot M_{\rm capt}=\frac{1}{4}
\left(\frac{R_{\rm capt}}{a_{\rm orb}}\right)^2 \dot M_{\rm w}=
\left(\frac{G\,M_{\rm x}}{a_{\rm orb}}\right)^2 \frac{\dot M_{\rm w}}
{V^4_{\rm w}}
= 5.7\,10^{-9}\,M_\odot {\rm y}^{-1}$
on average throughout the orbital period. Then the
luminosity due to accretion onto the WD surface is
$L_{\rm a}=\frac{G\,M_{\rm x}\, \dot M_{\rm a}}
{2\, R_{\rm x}}
\approx 10\,L_\odot$.
The optical and IUE data show that the hot component luminosity
$L_{\rm h} =
240 - 300 \,L_{\odot}$ during the bright state in 1981 -- 84, and that
the luminosity changed by a factor of 3.2 during the
transitions from low to bright state in 1981 and back in 1984
(Miko{\l}ajewski et al. 1990). On the other hand, in quiescence
the luminosity of the hot component was below
$1\,L_\odot$, that is much lower than the wind powered accretion
luminosity. Fig.\ref{orbita} demonstrates the
orbital dependence of the luminosity of the accretion
onto the WD surface in the
case of a spherical accretion of the giant's wind.
Both, the extremal values and sudden changes of the luminosity of the hot
component, as well as the jet ejection,
are problematic in a simple accretion model. The problems become
aggravated in connection with the quasi-periodic dips of brightness
(of relatively short-time duration, \si\ 3 months),
which were interpreted by Skopal et al. (1996)
as eclipses in a triple system, and with the ejection of clumps with
velocities up to 2500 km/s visible at least in the orbital plane
(Tomov et al. 1996).

   In the triple system the outbursts could be caused by episodic
enhanced stripping of the giant at eccentric 756$^d$ orbit of the
interior binary with following energetic accretion of the matter by
the compact companion at the outer orbit. However the quasi-periodic
 dips could
not be explained as eclipses in the triple system
(Mikkola \& Tanikawa 1998). Alternatively, the outbursts could be due
to instability of the accretion disk like that of dwarf novae
or X-ray transients. However,
there is no reliable evidence
of an accretion disk in \ch. Although the propeller is able to put in
the circulation of the disk.

   The problems may be overcome if one supposes  
that the compact component of the system is
magnetized enough to act like a propeller on ionized infalling
matter. In this case matter necessary for the forthcoming outburst
may be stored in a thick envelope, supported against the gravitation by
the propeller. Such an envelope will be lumpy and with a flat
distribution of density. When enough mass will be accumulated, the
envelope contracts the magnetosphere under the corotation radius and
accretion onto the star surface will follow, that is the accretion phase.
In the transition between the propeller
and accretion phases, through the corotation radius,
the envelope structure and the hot component
luminosity will abruptly change. The sudden
increase from $75\,L_\odot$ to $240\,L_\odot$ 
in the hot component luminosity in 1981
was probably such a catastrophic transition.
Therefore we suppose the maximum propeller luminosity
(when $R_{\rm m}=R_{\rm cor}$) is $L_{\rm p}=75\,L_\odot$
(the share of release of the gravitation energy of the gas
falling onto the magnetosphere of the corotation
radius $\sim 10^{10}$ cm, for a star's spin period
$P_{\rm x}= 500$\,s, is insignificant: $L_{\rm a} \approx 0.5\,L_\odot$).

   In the steady state on average properties of a propeller are
determined by the position of magnetosphere relative to the
corotation radius. The magnetosphere radius is approximately
determined by the competition between the magnetic field pressure and gas
pressure. In the case of $p_{\rm g}=p_{\rm ff}=
\frac{\dot M_{\rm cap} (2\,G\, M_{\rm x})^{1/2}}
{4\,\pi\, R_{\rm m}^{5/2}}$ the field strength will be of
order $B_{\rm s} \sim 10^7$ G (from Eq.\ref{bal}).
However in the case of the envelope around the
magnetosphere the idealization of free fall does not work and it is
necessary to calculate self-consistently the interaction
of the envelope with the magnetosphere to find $p_{\rm g} (R_{\rm m})$. 

In other way, the strength of the WD magnetic field
can be derived from Eq.\ref{lprop} for the maximum propeller luminosity,
i.e. when $R_{\rm m}=R_{\rm cor}$:
\begin{equation}
\label{bs}
B_{\rm s} \sim 6\,10^7
\left(\frac{L_{\rm p}(R_{\rm m}=R_{\rm cor})}
{100\,L_\odot}\right)^{1/2}
\left(\frac{P_{\rm x}}{10\,{\rm min}}\right)^{3/2}{\rm G}.
\end{equation}
This field is strong enough to be measured. This
would be possible only in the case of absence of the envelope. However
the hot blue spectrum is then overwhelmed by the giant, that makes
difficult the polarization measurements.

\begin{figure}
\centerline{\psfig{figure=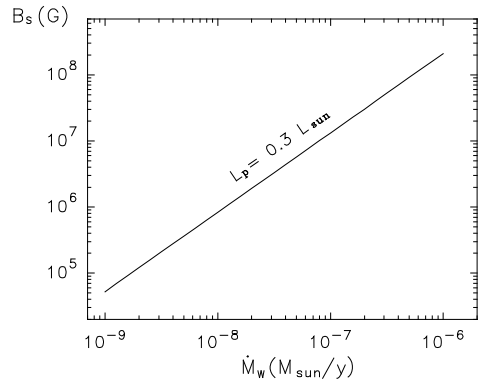,height=5cm}}
\caption{In the case of the magnetic propeller in CH\,Cyg
the permitted values of the mass loss rate
$\dot M_{\rm w}$ and the magnetic field strength
$B_{\rm s}$ are above the line
which corresponds to the propeller luminosity $L_{\rm p}=0.3\,L_\odot$.}
\label{field}
\end{figure}

In quiescence the hot component luminosity is lower than
$1\,L_\odot$, while the accretion onto the WD surface 
from the giant's wind would be more energetic (Fig.\ref{orbita}).
In a framework of the propeller model this is suppressed.
When the envelope is negligible
the magnetosphere radius may be approximately estimated from the balance
between the magnetic pressure at the upper boundary of the magnetosphere
and dynamical pressure of 
free falling gas. Eliminating $R_{\rm m}$ from
Eqs.~\ref{bal} and \ref{lprop}, we have a minimum
luminosity of the hot component produced by the interaction
of the rotating magnetic field with the wind:
\begin{eqnarray}
\label{lpropw}
L_{\rm p} \sim \frac{\eta^2 R_{\rm co}^3}{P_{\rm x}}
\frac{(8\,G\,M_{\rm x} \dot M_{\rm cap}^2)^{6/7}}
{(B_{\rm s} R_{\rm x}^3)^{10/7}}= \hspace{2.5cm}
\nonumber \\ \hspace{.5cm} 1.1\,\eta^2
\left(\frac{\dot M_{\rm cap}}{10^{-8}\, M_\odot\,{\rm y}^{-1}}\right)^{12/7}
\left(\frac{10^7\,{\rm G}}{B_{\rm s}}\right)^{10/7}\
L_\odot.
\end{eqnarray}
This equation defines a field of possible values of $B_{\rm
s}$ and $\dot M_{\rm w}$. In Fig.\ref{field} the permitted values
lie  above the line corresponding to the propeller luminosity
$0.3\,L_\odot$, that equals the minimum hot component luminosity
between outbursts (Murset et al. 1991), and to the parameters
$\eta=1$ and $\dot M_{\rm cap}/\dot M_{\rm w}=0.06$.
So, for $\dot
M_{\rm w}= 10^{-7}$\,\msy\ the field must be $\ga 1.5\,10^7$\,G.
This limit is in good agreement with the estimate of $B_{\rm s}$
made above on the base of the catastrophic transition, which is
evidence of
consistency of the propeller model with the observed
values of $L_{\rm h}$, $P_{\rm x}$ and $\dot M_{\rm w}$.

In a framework of the propeller model,
the dips of brightness of \ch\ may be caused by instability of
the accretion structure around the magnetosphere, whose energy release
strongly depends on radius:
$L_{\rm p} \propto R_{\rm m}^{-6}$ (Eq.\ref{lprop}). 
Possibly, the lumpy nebula around \ch, that
appeared after the notable dip in 1984,
was also caused by the propeller action. And it is evident
that the jet ejection was
immediately connected with the catastrophic transition between
the accretion and propeller phases, when coupling of the magnetic field and
gas is most effective (then the Alfven and corotation velocities are
equal).

\subsection{\mwc}
\begin{figure}
\centerline{\psfig{figure=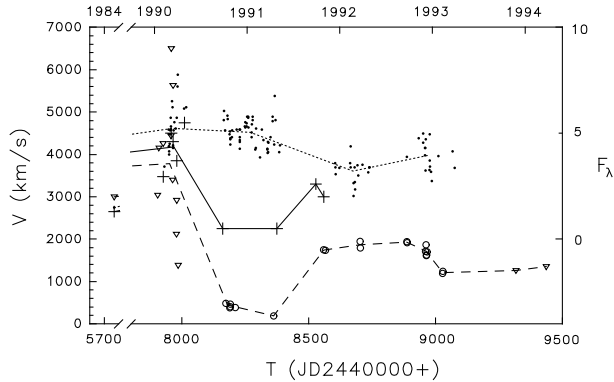,height=5cm}}
\caption{Time variations of MWC~560 activity. There are shown velocity 
of H$_\beta$ line core,
radiation fluxes in optics (dots) and 
in UV (crosses). The left axis --- velocity in \kms, the right ---
radiation flux in ${\rm 10^{-13}~erg\, cm^{-2}\,s^{-1}}$~\AA$^{-1}$.}
\label{mwccur}
\end{figure}

   \mwc\ is similar to \ch\ (Miko{\l}ajewski, Tomov \&
Miko{\l}ajewska 1997), but differs by the orbit inclination, which is
close to $0^\circ$. Therefore, the approaching jet in \mwc\ is
observed in blue-shifted absorption lines and sometimes the receding
jet is observed in red-shifted emission lines. The approaching jet
developed a maximum velocity of about 7000~\kms\ in the outburst in 1990
(Tomov, Kolev, Georgiev et al. 1990). It seems that the jets outflows
are inherent more or less permanently in polar directions (see
Fig.~\ref{mwccur}), at least from the first observation of the
blue-shifted high-velocity absorptions in 1943 (Merrill \& Burwell
1943). The gas temperature in the jets is 7000 -- 8000 K.
The blue hot
continuum of \mwc\ cannot be fitted either by an accretion disk
model or by a model of nuclear burning on the surface of the WD
(Panferov, Fabrika \& Tomov 1997). Moreover, these models cannot
account for the powerful wind and jets. The hot
component radiation originates in the photosphere of the slow
($\sim 100$ \kms) powerful ($10^{-6}$ \msy) wind with $T\approx
20000$ K and a radius of $4\,10^{11}$\,cm. The low-velocity
absorption and emission lines form in this wind. 

   From the similarity of both stars the high-velocity permanent polar
outflows probably may occur in \ch\ too. But they might be visible only
in outburst as this happened in the
biggest outburst in 1984 (Taylor, Seaquist \& Mattei 1986).
The
permanent jets suggest a permanent jet engine to exist
in these stars independent of
details of the accretion structure (is it an envelope, or an accretion
disk, or a ring), which strongly supports a propeller  as the engine.

   \mwc\ is one of a few SSs having flickering. Miko{\l}ajewski et
al. (1998) reported a huge amplitude flickering of 0.7 magnitude. In
1992 we observed a rare burst in the H$_\beta$ line with an amplitude
of 3 for half an hour, which resembles quasi-regular bursts in AE\,Aqr.
Besides there are semicoherent 22 min variations (Dobrzycka,
Kenyon \& Milone 1996). The orbital inclination of \mwc\ close to
$0^\circ$ gives us a happy chance to be in the cone of the jet and UV
radiation from the deep of the photosphere. This explains why the hot
component radiation is more hard and variable in \mwc\ than in \ch.
Also, the correlation of the UV flux and the jet velocity
(Fig.~\ref{mwccur}) may be due to the effect of variable optical
thickness of the jet.

The observed luminosity of \mwc\ is much lower than the critical
Eddington luminosity. Therefore the radiation mechanisms cannot account
for the wind and the jets.
We suppose that they
are accelerated by a propeller action of the magnetic WD.
Then the outburst in 1990 could correspond to the catastrophic
transition of the propeller, which means that its luminosity was
$L_{\rm p}\sim 10^3\,L_\odot$ when $R_{\rm m}=R_{\rm co}$.
From Eq.~\ref{bs} this gives a magnetic field strength
$B_{\rm s} \sim 6\,10^8$\,G for the supposed WD's spin period
${\rm P_x = 22}$~min.

\section{Summary}
The propeller model allows both the nature of the
hot component of the binary,
and the origin of the jets to be explained
in \ch\ and \mwc. The observational
signatures of a propeller, as inferred from \ch\ and \mwc, may be
as follow:
\begin{enumerate}
\item Erratic variability, when the magnetosphere radius is close or
under the corotation radius.
\item Dips of brightness at the times of outbursts, which may be related
to a transition accretor -- propeller or may be caused by the instability
of the accretion structure at the propeller phase.
\item Wind outflow and jet ejection, which have a maximum rate near the
catastrophic transition.
\item Hot component radiation originates in highly inhomogeneous envelope
around WD.
\end{enumerate}

The propeller model implies a magnetic field of WD in \ch\ and \mwc\
to be of order $10^8$~G. Measurement of such fields
in polarization observations would be a straight way to verify
the model. However this seems to be a very involved problem:
the star surface and the magnetosphere are
hidden from the observer under the photosphere of the outflowing wind.

In this paper we suggest only a scheme of the propeller model.
It is believed that gas outflows form around the propeller
in some sectors,
or nonstationary gas ejection occurs when accretion is replaced by outflow.
In some propeller scenarios envelopes or disks may form round
the magnetosphere. But to define
specifically the picture of accretion and outflow has been
impossible so far.
So further investigations of the accretion structure
in \ch\ and \mwc\ are important.

\begin{acknowledgements} This work was supported by the Polish
KBN Research Grant 2PO3D\,019\,17.
\end{acknowledgements}

\end{document}